\documentstyle[12pt]{article}

\global\arraycolsep=1pt
\begin{document}

\hfill    SISSA/ISAS 67/94/EP

\hfil     hepth@xxx/9405174

\hfill    IFUM 469/FT

\hfill    May, 1994

\begin{center}
\vspace{10pt}
{\large \bf
CONSTRAINED TOPOLOGICAL FIELD THEORY
\, \footnotemark
\footnotetext{Partially supported by EEC,
Science Project SC1$^*$-CT92-0789.}
}
\vspace{10pt}

{\sl Damiano Anselmi, Pietro Fr\'e}

\vspace{4pt}

International School for Advanced Studies (ISAS), via Beirut 2-4,
I-34100 Trieste, Italia\\
and Istituto Nazionale di Fisica Nucleare (INFN) - Sezione di Trieste,
Trieste, Italia\\

\vspace{8pt}

{\sl Luciano Girardello and Paolo Soriani}

\vspace{4pt}

Dipartimento di Fisica, Universit\`a di Milano, via Celoria 6,
I-20133 Milano, Italia\\
and Istituto Nazionale di Fisica Nucleare (INFN) - Sezione di Milano,
Milano, Italia\\
\end{center}

\vspace{8pt}

\begin{center}
{\bf Abstract}
\end{center}

\vspace{4pt}

\noindent

We derive a model of constrained topological gravity,
a theory recently introduced by us
through the twist of N=2 Liouville theory, starting from the
general BRST algebra and imposing the moduli space constraint as a
gauge fixing. To do this, it is necessary to introduce
a formalism that allows a careful treatment
of the global and the
local degrees of freedom of the fields.
Surprisingly, the moduli space
constraint arises from the simplest and most natural gauge-fermion
({\sl antighost} $\times$ {\sl Lagrange multiplier}), confirming
the previous results.
The simplified technical set-up provides a deeper understanding for
constrained topological gravity and a convenient
framework for future investigations, like the matter coupling and
the analysis of
the effects of the constraint on the holomorphic anomaly.
\vfill
\eject

In a recent paper \cite{preprint}, we analysed the topological
twist of N=2 supergravity in two dimensions and revealed
some new features with respect to the known models of
topological gravity. The key point
(that, to our knowledge, was not previously noticed in the literature)
is that the gravitini must be $U(1)$ charged with respect to the
graviphoton in order to close off-shell the
supersymmetry transformations\footnotemark
\footnotetext{The corresponding supersymmetry algebra was called
by us {\sl charged Poincar\`e} superalgebra.}. This fact has crucial
effects: the graviphoton $A$ appears, after the twist, in the
BRST variation of some antighost and can be interpreted as a BRST
Lagrange multiplier; moreover, the $U(1)$ current is nontrivial and
its vanishing projects the moduli space ${\cal
M}_g$ of the genus $g$ Riemann surfaces onto a homology cycle
${\cal V}_g\in H_{2g-3}({\cal M}_g)$ of codimension $g$.

Any topological field theory projects the functional integral onto
the moduli space of some {\sl
instantons}. Consequently,
the Riemann surfaces lying in ${\cal V}_g$ are
to be named {\sl gravitational instantons in two dimensions}.

In the present paper, we construct a model of constrained topological
gravity independently of any topological twist, that however captures
the main suggestion springing from the twist of the N=2 theory. For
the details of the twisted model as well as
for many other technical points
just alluded to in the present letter, the reader is referred to ref.\
\cite{preprint}. Here, in any case, we also trace back the
match with the model of ref.\ \cite{preprint}. We get a theory whose
formal structure is essentially
the same as in the Verlinde and Verlinde model
\cite{verlindesquare}, but such that the correlation
functions are calculated on ${\cal V}_g$ instead of ${\cal M}_g$.

As discussed in detail in ref.\ \cite{preprint},
constrained topological gravity
is described by the gauge-free BRST algebra of $SL(2,{\bf R})$,
the same as in the Verlinde and Verlinde model \cite{verlindesquare},
\begin{eqnarray}
se^\pm&=&\psi^\pm-\nabla c^\pm\mp c_0 e^\pm,\nonumber\\
s\psi^\pm&=&-\nabla\gamma^\pm\mp c_0 \psi^\pm\pm\gamma_0e^\pm
\pm\psi_0c^\pm,\nonumber\\
s\gamma^\pm&=&\mp c_0\gamma^\pm\pm\gamma_0c^\pm,\nonumber\\
sc^\pm&=&\gamma^\pm\mp c_0c^\pm,\nonumber\\
\matrix{
s\omega=\psi_0-dc_0,}&&\matrix{
s\psi_0=-d\gamma_0,\quad &\quad
sc_0=\gamma_0,\quad &\quad s\gamma_0=0,}
\end{eqnarray}
but the gauge-fixing BRST algebra ${\cal B}_{gauge-fixing}$
is enriched with an antighost one-form
$\bar \psi$, a Lagrange multiplier one form $A$ and the respective
gauge ghosts, $\gamma$ and $c$,
\begin{equation}
\matrix{
s\bar\psi=A+d\gamma,\quad &\quad sA=dc,\quad &\quad s\gamma=c,
\quad &\quad sc=0.}
\label{bgf}
\end{equation}
The possibility of introducing a BRST constraint on the moduli space is due to
the fact that ${\cal B}_{gauge-fixing}$
contains fields ($\bar \psi$ and $A$) that
possess global degrees of freedom: instead of enlarging the moduli space,
they {\sl reduce} it.

Due to this, it is necessary to develop a formalism that
permits to deal conveniently
with the global degrees of freedom, together with
the local ones\footnotemark\footnotetext{For this problem
see also \cite{imbimbo}.}.
Let us introduce fiducial fields $\hat e^\pm$,
$\hat \psi^\pm$, $\skew5\hat{\bar\psi}$ and $\hat A$ satisfying the
``purely topological'' BRST algebr\ae\
\begin{equation}
\matrix{s\hat e^\pm=\hat\psi^\pm,\quad &\quad s\hat\psi^\pm=0,
\quad &\quad s\skew5\hat{\bar\psi}=\hat A,\quad &\quad s\hat A=0.}
\label{sgfil}
\end{equation}
The hatted fields will represent the purely global degrees of freedom of the
corresponding unhatted fields. For example, $\hat e^\pm$ represents
the conformal class of the metric represented by $e^\pm$.

To clarify this point, it is useful to write
down the gauge-fixings of diffeomorphisms and
Lorentz rotations together with their fermionic counterparts,
\begin{eqnarray}
e^+\wedge \hat e^+=0,\quad &\quad e^-\wedge \hat e^-=0,\quad &\quad
e^+\wedge \hat e^-+e^-\wedge \hat e^+=0,\nonumber\\
\psi^+\hat e^+=\hat \psi^+e^+,\quad &\quad
\psi^-\hat e^-=\hat \psi^-e^-,\quad &\quad
\psi^+\hat e^-+\psi^-\hat e^+=\hat \psi^+e^-+\hat \psi^-e^+.
\end{eqnarray}
We can solve these gauge-fixing conditions by writing
\begin{equation}
e^\pm={\rm e}^\varphi \hat e^\pm,\quad\quad
\psi^\pm={\rm e}^\varphi (\eta \hat e^\pm+\hat \psi^\pm),
\end{equation}
$\varphi$ being the Liouville field and $\eta$ its superpartner.
We see that
$\hat e^\pm$ represent the differentials $dz$ and $d\bar z$. Moreover,
writing
$\hat\psi^\pm=\hat \psi^\pm_+\hat e^++\hat \psi^\pm_- \hat e^-$, we
are lead to identify
$\hat\psi^+_-$ and $\hat\psi^-_+$
with $\hat m_i\mu^{iz}_{\bar z}$ and $\hat{\bar m}_i {\bar \mu}^{i\bar z}_z$,
respectively. Here $\hat m_i$ and $\hat{\bar m}_i$,
$i=1,\ldots ,3g-3$ are the supermoduli,
while $\mu^{iz}_{\bar z}$ and ${\bar \mu}^{i\bar z}_z$ are
the Beltrami differentials. On the other hand, writing
$\skew5\hat{\bar\psi}=\skew5\hat{\bar\psi}_+\hat e^++\skew5\hat{\bar\psi}_-\hat
e^-$
and $\hat A=\hat A_+\hat e^++\hat A_-\hat e^-$, we can identify
$\skew5\hat{\bar\psi}_+$ and $\skew5\hat{\bar\psi}_-$
with $\hat \nu_j \omega^j_z$ and
$\hat{\bar \nu}_j{\bar \omega^j}_{\bar z}$, while
$\hat A_+$ and $\hat A_-$ can be identified with
$\nu_j\omega^j_z$ and ${\bar \nu_j}{\bar \omega}^j_{\bar z}$.
Here $\omega^j_z$ is a basis of $g$ holomorphic differentials,
$g$ being the genus of the Riemann surface, and ${\bar \omega}^j_{\bar z}$
are their complex conjugates.
Finally, $\nu_j$ and $\bar \nu_j$ are the global degrees of freedom of
the
$U(1)$ gauge field $A$,
while $\hat \nu_j$ and $\hat{\bar \nu}_j$ are those of $\bar \psi$.
Formula (\ref{sgfil}) should be compared with formula (9.1) of
\cite{preprint}.

The functional measure $d\mu$ contains the integration $d\hat\mu$
over the hatted fields,
\begin{equation}
d\hat\mu=d\hat e^+d\hat e^-d\hat \psi^+ d\hat \psi^-d\hat A
d\skew5\hat{\bar \psi},
\end{equation}
where
$d\hat e^+d\hat e^-$ is the integration over the moduli space ${\cal
M}_g$ of genus $g$ Riemann surfaces $\Sigma_g$, while
$d\hat A$ is the integration over the moduli space of $U(1)$ flat
connections,
which is the Jacobian variety ${\bf C}^g/({\bf Z}^g+\Omega
{\bf Z}^g)$, $\Omega$ being the period matrix of $\Sigma_g$.
$d\hat \psi^+ d\hat \psi^-$ and $d\skew5\hat{\bar\psi}$
are the integrations over the supermoduli.

The above identifications between hatted fields and moduli
will be made only in the final expressions:
in the intermediate steps it is convenient to retain the hatted fields,
in order to avoid concepts like the
``field dependent points'' $z$ and $\bar z$, that are unusual in
quantum field theory.

Let $\Omega^{(i)}=\Omega^{(i)}_+\hat e^++\Omega^{(i)}_-\hat e^-$,
$i=1,2$, be two one-forms such that $s\Omega^{(1)}=\Omega^{(2)}$.
Then the formul\ae\ for $s\Omega^{(1)}_+$ and $s\Omega^{(1)}_-$
read
\begin{eqnarray}
s\Omega^{(1)}_+&=&\Omega^{(2)}_+-\hat \psi^+_+\Omega_+^{(1)}
-\hat \psi^-_+\Omega_-^{(1)},\nonumber\\
s\Omega^{(1)}_-&=&\Omega^{(2)}_--\hat \psi^+_-\Omega_+^{(1)}
-\hat \psi^-_-\Omega_-^{(1)}.
\label{poj}
\end{eqnarray}
These expressions will be crucial when constraining the moduli space.
Notice that even when $\Omega^{(2)}=0$, $s\Omega^{(1)}_\pm$ are
nonzero.

Another important point concerns the gauge-fixing of the local $U(1)$
gauge symmetries. In section VII of
ref.\ \cite{preprint} we invented a suitable trick in order
to reach a complete chiral factorization between left
and right moving sectors,
at least in the limit when the cosmological constant
tends to zero.
Here, we need to use the same trick
twice, once for $\bar\psi$ and once for $A$. So, let us introduce
two trivial BRST systems $\{\bar\Gamma,\zeta\}$ and $\{\xi,c^\prime\}$, with
\begin{eqnarray}
s\bar \Gamma=\zeta,\quad &&\quad s\zeta=0,\nonumber\\
s\xi=c^\prime,\quad &&\quad sc^\prime=0.
\end{eqnarray}
The gauge-fixings of the $\gamma$ and $c$ symmetries, as well as the
above two trivial symmetries are
\begin{equation}
\bar\psi_\pm=\skew5\hat{\bar\psi}_\pm\mp\nabla_\pm\bar \Gamma,
\quad\quad
A_\pm=\hat A_\pm\mp\nabla_\pm\xi,
\end{equation}
where $\nabla_\pm$ are such that the exterior derivative $d$ is
$\hat e^+\nabla_++\hat e^-\nabla_-$.

Finally, we need to fix the topological symmetries and this can
be achieved by setting
\begin{equation}
T^\pm=0,\quad\quad R=a^2e^+ e^-,\quad \quad d\bar\psi=0,
\end{equation}
$T^\pm=de^\pm\pm\omega\wedge e^\pm$ being the torsions.
The condition $d\bar\psi=0$ and its BRST variation $dA=0$
do not depend on the global degrees of freedom $\nu$ and $\hat \nu$.
The gauge-fixing for them (moduli space constraint) will be treated in detail
later on.

With obvious notation, the gauge fermions for the gauge-fixings that
we have so far introduced are
\begin{eqnarray}
\Psi_1&=&\bar \pi_\pm T^\pm+\bar \pi(R-a^2e^+e^-),\nonumber\\
\Psi_2&=&\chi d\bar \psi,\nonumber\\
\Psi_3&=&b_{++}e^+\hat e^++b_{--} e^-\hat e^-+b_{+-}
(e^+\hat e^-+e^-\hat e^+)+
\beta_{++}(\psi^+\hat e^+-\hat \psi^+e^+)\nonumber\\&&
+\beta_{--}
(\psi^-\hat e^--\hat \psi^-e^-)+\beta_{+-}
(\psi^+\hat e^-+\psi^-\hat e^+-\hat \psi^+e^--\hat \psi^-e^+),
\nonumber\\
\Psi_4&=&\beta_+(\bar\psi_--\skew5\hat{\bar\psi}_--\nabla_-\bar \Gamma)
\hat e^+\hat e^-+
b_+(A_--\hat A_--\nabla_-\xi)\hat e^+\hat e^-\nonumber\\&&
+\beta_-(\bar\psi_+-\skew5\hat{\bar\psi}_++\nabla_+\bar \Gamma)
\hat e^+\hat e^-+
b_-(A_+-\hat A_++\nabla_+\xi)\hat e^+\hat e^-.
\label{gfs}
\end{eqnarray}

Let $\pi=s\bar \pi$ and $\lambda=s\chi$.
Focusing on the local degrees of freedom, the Lagrangian ${\cal L}=
-s\sum_{i=1}^4
\Psi_i$ turns out to be, after some simple field redefinitions similar
to those discussed in section VII of ref.\ \cite{preprint} and
in the limit $a^2\rightarrow 0$,
\begin{eqnarray}
{\cal L}&=&\pi\partial_z\partial_{\bar z}\varphi+
\chi\partial_z\partial_{\bar z}\xi-
\bar \pi\partial_z\partial_{\bar z}\eta+\lambda\partial_z\partial_{\bar z}
\bar \Gamma\nonumber\\
&&-b_{zz}\partial_{\bar z} c^z-\beta_{zz}\partial_{\bar z}\gamma^z
-\beta_z\partial_{\bar z}\gamma-b_z\partial_{\bar z}c
\nonumber\\&&
+b_{\bar z\bar z}\partial_{z} c^{\bar z}+\beta_{\bar z\bar z}\partial_{z}
\gamma^{\bar z}-\beta_{\bar z}\partial_{z}\bar \gamma-b_{\bar z}
\partial_{z}\bar c.
\end{eqnarray}
Notice that there are second order fermions, differently from
the model of ref.\ \cite{preprint}.

The BRST charge ${\cal Q}$
can be easily found with the method of
ref.\ \cite{preprint}, i.e.\ by means of a
local BRST variation of the Lagrangian. One can then
write
\begin{equation}
{\cal Q}={\cal Q}_s+{\cal Q}_v,
\end{equation}
where ${\cal Q}_s^2={\cal Q}_v^2=\{{\cal Q}_s,{\cal Q}_v\}=0$. Explicitly,
\begin{equation}
{\cal Q}_s=
\oint -\partial_z\pi\eta-\lambda\partial_z\xi+b_{zz}\gamma^z
-\beta_z c,
\end{equation}
while ${\cal Q}_v$ is the same as in the Verlinde and Verlinde model
plus $-\partial_z\chi c-\partial_z\lambda\gamma$. To recover the correct
energy momentum tensor, one has to perform the following redefinitions
(left moving part)
\begin{eqnarray}
\quad &b_{zz}\rightarrow b_{zz}-b_z\partial_z\gamma+\partial_z\chi\partial_z
\bar \Gamma,\quad &\quad
\beta_z \rightarrow \beta_z+\partial_z (b_z c^z),
\nonumber\\
\quad &\xi\rightarrow \xi-\partial_z\bar \Gamma c^z,\quad &\quad
\lambda\rightarrow \lambda+\partial_z\chi c^z,\nonumber\\
\quad &c\rightarrow c +\partial_z\gamma c^z.\quad &\quad
\label{redef}
\end{eqnarray}
The operator product expansions  among the fields are left unchanged
(and so the form of the Lagrangian). The final BRST charge ${\cal Q}$ is
${\cal Q}_s+{\cal Q}_v$ with ${\cal Q}_s$ as before and
\begin{equation}
{\cal Q}_v=\oint -c^z{\cal T}_{zz}-{1\over 2}
\gamma^z{\cal G}_{zz}-{1\over 2}c{\cal J}_z^\prime+{1\over 2}\gamma
{\cal G}_z^\prime.
\end{equation}
One has\footnotemark\footnotetext{We have normalized the currents in
such a way that $\{Q_s, G_{zz}\}=2T_{zz}.$}
\begin{eqnarray}
{\cal T}_{zz}&=&T^{grav}_{zz}+{1\over 2}T^{gh}_{zz},\quad\quad
{\cal G}_{zz}=G^{grav}_{zz}+{1\over 2}G^{gh}_{zz},\nonumber\\
T^{grav}_{zz}&=&T^{grav}_{(1)zz}+T^{grav}_{(2)zz},
\,\,\quad\quad
G^{grav}_{zz}=G^{grav}_{(1)zz}+G^{grav}_{(2)zz},\nonumber\\
T^{gh}_{zz}&=&T^{gh}_{(1)zz}+T^{gh}_{(2)zz},\quad\quad\,\,\,
G_{zz}^{gh}=G_{(1)zz}^{gh}+G_{(2)zz}^{gh},
\end{eqnarray}
and
\begin{eqnarray}
T^{grav}_{(1)zz}&=&\partial_z\pi\partial_z\varphi-{1\over 2}\partial_z^2
\pi -\partial_z\bar \pi\partial_z\eta,
\quad\quad\quad T^{grav}_{(2)zz}=\partial_z\chi\partial_z\xi +\partial_z
\lambda\partial_z \bar \Gamma,\nonumber\\
G^{grav}_{(1)zz}&=&\partial_z^2\bar \pi -2\partial_z\bar \pi\partial_z\varphi,
\quad\quad\quad G^{grav}_{(2)zz}=
-2\partial_z\chi\partial_z\bar \Gamma,\nonumber\\
T^{gh}_{(1)zz}&=&2b_{zz}\partial_zc^z+\partial_zb_{zz}c^z
+2\beta_{zz}\partial_z\gamma^z +\partial_z\beta_{zz}\gamma^z,
\quad\quad\quad T^{gh}_{(2)zz}=
\beta_z\partial_z\gamma +b_z\partial_zc,\nonumber\\
G_{(1)zz}^{gh}&=&-4\beta_{zz}\partial_zc^z-2\partial_z\beta_{zz}c^z,
\quad\quad G_{(2)zz}^{gh}=2b_z\partial_z\gamma.
\label{topoal}
\end{eqnarray}
On the other hand, the $U(1)$ current $J^\prime_z=2\{{\cal Q}_v,b_z\}$
and the associated supercurrent $G^\prime_z=-2[{\cal Q}_v,\beta_z]$ are
trivial:
\begin{eqnarray}
J^\prime_z=2\partial_z\chi+2\partial_z(b_zc^z),\quad&\,\,
{\cal J}^\prime_z=2\partial_z\chi+\partial_z(b_zc^z),\nonumber\\
G^\prime_z=-2\partial_z\lambda+2\partial_z(\beta_zc^z)+2
\partial_z(b_z\gamma^z),\quad &\,\,
{\cal G}^\prime_z=-2\partial_z\lambda+\partial_z(\beta_zc^z)+
\partial_z(b_z\gamma^z),
\end{eqnarray}
in the sense that $\int\bar \omega^j_{\bar z}J^\prime_z=0$
and $\int\bar \omega^j_{\bar z}G^\prime_z=0$, $\forall j$.
Recall that in
ref.\ \cite{preprint} it was precisely the nontriviality of
$\int\bar \omega^j_{\bar z}J_z$ that was responsible
of the constraint on the
moduli space. This was due to the outlined
fact that the gravitini need to be $U(1)$ charged
in order to close the N=2 supersymmetry
transformations {\sl off-shell}.
In the present model, however, we have not dealt so far with the moduli space
constraint and no field is $U(1)$ charged. Indeed,
the present approach is ``constructive'', in the sense
that we are not getting the topological theory
from an already formulated independent
model (like N=2 Liouville theory). This means that
the moduli space constraint has to be introduced ``by hand''.

As a matter of fact, the topological algebra is not closed by
$G^\prime_z$ and $J_z^\prime$, but by the {\sl topological current}
$G_z$, such that ${\cal Q}_s=\oint G_z$, and by the ghost current $J_z$,
\begin{eqnarray}
G_z&=&G^{grav}_{(1)z}+G^{grav}_{(2)z}+G^{gh}_{(1)z}+G^{gh}_{(2)z},\quad\quad
J_z=J^{grav}_{(1)z}+J^{grav}_{(2)z}+J^{gh}_{(1)z}+J^{gh}_{(2)z},\nonumber\\
G^{grav}_{(1)z}&=&-\partial_z\pi\eta,\quad
G^{grav}_{(2)z}=-\lambda\partial_z\xi,\quad
G^{gh}_{(1)z}=b_{zz}\gamma^z,\quad
G^{gh}_{(2)z}=-\beta_z c,\nonumber\\
J^{grav}_{(1)z}&=&-{1\over 2}\partial_z\pi + \eta\partial_z\bar \pi,
\quad J^{grav}_{(2)z}=
\lambda\partial_z\bar\Gamma,\quad
J^{gh}_{(1)z}=b_{zz}c^z+2\beta_{zz}\gamma^z,\quad
J^{gh}_{(2)z}=b_z c.\nonumber\\
\end{eqnarray}
This is more similar to what happens in the Verlinde and Verlinde
model. Indeed, the topological algebra $T_{zz}$-$G_{zz}$-$G_z$-$J_z$
is the tensor product of the Verlinde and Verlinde one
(denoted by the subscript $1$), and the {\sl constraining}
topological algebra, corresponding to equation
(\ref{bgf}) and denoted by
the subscript $2$.
The Verlinde and Verlinde topological algebra can be
untwisted
according to $T_{zz}\rightarrow T_{zz}-{1\over 2}\partial_zJ_z$,
to give an N=2 superconformal algebra with central charge
$c_{1}=c_{1}^{grav}+c_{1}^{gh}$, $c_{1}^{grav}=3$,
$c_{1}^{gh}=-9$, while
the constraining topological algebra can be
untwisted
to give an N=2 superconformal algebra with central charge
$c_{2}=c_{2}^{grav}+c_{2}^{gh}$, $c_{2}^{grav}=3$,
$c_{2}^{gh}=3$. Thus the total topological algebra has central
charge $c=c^{grav}+c^{gh}$, $c^{grav}=6$,
$c^{gh}=-6$, as in ref.\ \cite{preprint}.
Nevertheless, the above representation of the topological algebra
is {\sl different} from the one of ref.\ \cite{preprint}.
A map between the present conformal theory and the one
of ref.\ \cite{preprint} is easily derived as follows\footnotemark
\footnotetext{For this pourpose,
it is convenient to introduce first order fermions $\bar \pi_z=
\partial_z\bar \pi$ and $\bar \Gamma_z=\partial_z\bar \Gamma$ (and their
complex conjugates).}.
Following a procedure similar to the one of section VIII
of \cite{preprint}, we can write (see also \cite{eguchietal})
\begin{equation}
{\cal Q}_v=\{{\cal Q}_s,[{\cal Q}_v,S]\},\quad
S=\oint b_z\gamma-\beta_{zz}c^z,\quad S^2=0,\quad
U_1={\rm e}^{[{\cal Q}_v,S]},\quad U_1{\cal Q}U_1^{-1}={\cal Q}_s.
\end{equation}
Now, the {\sl topological charge} ${\cal Q}_s$ is the same as in the
theory of ref.\ \cite{preprint}
and there exists an operator $U_2$
(see section VIII of \cite{preprint})
such that $U_2{\cal Q}^\prime U_2^{-1}={\cal Q}_s$,
${\cal Q}^\prime={\cal Q}_s+{\cal Q}_v^\prime$ denoting the total BRST
charge of the topological model of ref.\ \cite{preprint}.
Then the operator $U=U_2^{-1}U_1$ maps between the conformal field
theories corresponding to the two models of constrained topological
gravity:
\begin{equation}
{\cal Q}^\prime=U{\cal Q}U^{-1}.
\end{equation}
The ``singular'' character of $U_2$ (the field redefinitions
contain negative powers of $1-\gamma$) is thus explained by the fact
that the $U(1)$ currents are different in the two cases and
the moduli space constraint is imposed in a different way.

We now discuss the constraint on the moduli space. It is
easy to see that the Lagrangian ${\cal L}=\sum_{i=1}^4-s\Psi_i$
is independent of the global degrees of freedom $\nu$ and $\hat \nu$.
Indeed, from (\ref{poj}) and (\ref{gfs}) it follows that ${\cal L}_4=-s\Psi_4$
only depends on the differences $A_\pm-\hat A_\pm=\mp\nabla_\pm\xi$ and
$\bar\psi_\pm-\skew5\hat{\bar\psi}_\pm=\mp\nabla_\pm\bar \Gamma$ and never
on $A_\pm$, $\bar\psi_\pm$, $\hat A_\pm$, $\skew5\hat{\bar\psi}_\pm$,
separately. On the other hand, ${\cal L}_2=-s\Psi_2$ contains
$d\bar\psi$ and $dA$, which are the same as $d(\bar\psi-\skew5\hat{\bar\psi})$
and $d(A-\hat A)$, if we take into account that
$d\skew5\hat{\bar\psi}=0$ and $d\hat A=0$.

Thus the $\nu$-$\hat\nu$ dependence is completely confined
to a fifth gauge fermion $\Psi_5$, by means of which we now
impose the moduli space constraint.
We have two possibilities that are related to two
different descriptions given in ref.\ \cite{preprint}.

The first possibility is represented by a rather natural gauge-fermion:
{\sl antighost} $\times$ {\sl Lagrange multiplier}; precisely
\begin{equation}
\Psi_5={1\over 2}
(\skew5\hat{\bar\psi}_+\hat A_-+\skew5\hat{\bar\psi}_-\hat A_+)\hat e^+\hat
e^-.
\label{gf51}
\end{equation}
Then, we have, using (\ref{sgfil}) and (\ref{poj}),
\begin{eqnarray}
{\cal L}_5&=&s\Psi_5=(\hat A_+\hat A_--\hat \psi^-_+\skew5\hat{\bar\psi}_-\hat
A_-
-\hat\psi^+_-\skew5\hat{\bar\psi}_+\hat A_+)\hat e^+\hat e^-\nonumber\\
&=&\nu_j\bar\nu^k \int_{\Sigma_g}\omega^j_z\bar\omega^k_{\bar z}d^2z
-\hat{\bar m}_i\hat{\bar \nu}_j\bar\nu^k\int_{\Sigma_g}\bar\mu^{i\bar z}_z
\bar\omega^j_{\bar z}\bar\omega^k_{\bar z}d^2z-
\hat m_i\hat\nu_j\nu^k\int_{\Sigma_g}\mu^{i z}_{\bar z}
\omega^j_{z}\omega^k_{z}d^2z.
\end{eqnarray}
Now we want to perform the $\nu$-$\hat \nu$ integration.
First of all, we notice that the $\nu$ integration can be performed
over all ${\bf C}^g$ instead of ${\bf C}^g/({\bf Z}^g+\Omega
{\bf Z}^g)$. Indeed, the restriction to ${\bf C}^g/({\bf Z}^g+\Omega
{\bf Z}^g)$ is due to the invariance with respect to the $U(1)$ gauge
transformations that are not continuously deformable to the identity.
However, such invariance is explicitly broken by $\Psi_5$, since the
$\hat A_+\hat A_-$ term is a kind of mass term for the $U(1)$
connection. When there is a gauge-invariance, one can break it
either by solving a
certain gauge-fixing condition or by introducing a
corresponding gauge-fermion in the action. The first possibility is
not practicable, in general, since solving a gauge-fixing condition
usually requires to invert derivative operators.
In the present case, however, the two possibilities are equally practicable,
but the second one is more convenient.
The BRST equivalence of the two possibilities can be proved by using
a stretching argument
like the one of ref.\ \cite{preprint} and convert the integration over
${\bf C}^g/({\bf Z}^g+\Omega
{\bf Z}^g)$ to the integration over the full ${\bf C}^g$.

Using the properties (see \cite{dhoker} for example)
\begin{equation}
\int_{\Sigma_g}\omega^j_z\bar\omega^k_{\bar z}d^2z=(\bar\Omega-\Omega)^{jk},
\quad\,\,
\int_{\Sigma_g}\mu^{i z}_{\bar z}
\omega^j_{z}\omega^k_{z}d^2z=i{\partial \Omega^{jk}\over
\partial m_i},\quad\,\,
\int_{\Sigma_g}\bar\mu^{i\bar z}_z
\bar\omega^j_{\bar z}\bar\omega^k_{\bar z}d^2z=-i{\partial\bar\Omega^{jk}
\over\partial\bar m_i},
\end{equation}
and using the well known formul\ae\ for a superdeterminant, we get
\begin{equation}
\int\prod_{j=1}^gd\nu_jd\bar\nu_jd\hat\nu_jd\hat{\bar \nu}_j\,\,
{\rm e}^{\int_{\Sigma_g}{\cal L}_5}=
\det\left({1\over \Omega-\bar\Omega}\bar\partial\bar\Omega
{1\over \Omega-\bar \Omega}\partial\Omega\right),
\end{equation}
where $\partial=\hat m_i{\partial\over\partial m_i}$ and a suitable
normalization factor has been introduced. This
is precisely the top Chern class $c_g({\cal E}_{hol})$
of the Hodge vector bundle ${\cal E}_{hol}\rightarrow {\cal M}_g$
whose sections are the holomorphic differentials. This representation
of $c_g({\cal E}_{hol})$ is easily obtained (see section IX of
ref.\ \cite{preprint})
by choosing the imaginary part of the period matrix as fiber metric.
It is amazing to notice that this result follows naturally from the
simplest
gauge-fermion that comes to one's mind, i.e.\ (\ref{gf51}). In some sense,
we can still say that the constraint comes automatically and is not
imposed ``by hand'', since at first sight there is no gauge-fixing condition
in (\ref{gf51}).

Now, let us discuss a second possibility, which better ``simulates''
what one gets automatically by twisting the N=2 Liouville theory.
Again, the form of the gauge-fermion is quite typical, namely
{\sl antighost} $\times$ {\sl gauge-condition}. This requires, however, that
we know {\sl a priori} what condition to impose.
Let ${\cal S}={\cal S}_zdz$ be a section of ${\cal E}_{hol}$.
As discussed in \cite{preprint}, we can project onto the Poincar\`e dual
of $c_g({\cal E}_{hol})$ by requiring the vanishing of
$a_j\equiv\int_{\Sigma_g}
\bar\omega^j_{\bar z}{\cal S}_z d^2z$ $\forall j$ \cite{griffithsharris}.
This is achieved by choosing
\begin{equation}
\Psi_5=\skew5\hat{\bar\psi}_-\hat e^-{\cal S}+\skew5\hat{\bar\psi}_+\hat e^+
\bar {\cal S}.
\label{gf52}
\end{equation}
Then, we get
\begin{eqnarray}
{\cal L}_5&=&s\Psi_5=(\hat A_--\hat\psi^+_-\skew5\hat{\bar\psi}_+)\hat e^-
{\cal S}
+\hat \psi^-_+\skew5\hat{\bar\psi}_-\hat e^+{\cal S}+\skew5\hat{\bar\psi}_-\hat
e^-
s{\cal S}+{\rm h.c.}\nonumber\\&=&
(\nu_j\bar a_k-\bar \nu_j a_k+\hat \nu_js\bar a_k-
\hat{\bar \nu}_jsa_k)(\bar\Omega-\Omega)^{jk}+
{\cal R},
\end{eqnarray}
where ${\cal R}$ is an addend made of terms
proportional to $a_j$ or $\bar a_j$ and independent
of $\nu$-$\bar \nu$. The integration over $\nu$-$\bar \nu$
gives delta functions that permit to neglect ${\cal R}$. At the end,
noticing that $sa_j$ can be replaced by $da_j$, $d$ being the
exterior derivative on the moduli space, we get
\begin{equation}
\int\prod_{j=1}^gd\nu_jd\bar\nu_jd\hat\nu_jd\hat{\bar \nu}_j\,\,
{\rm e}^{\int_{\Sigma_g}{\cal L}_5}=\left|\prod_{j=1}^g\delta(a_j)\,da_j
\right|^2,
\end{equation}
which is the representation of $c_g({\cal E}_{hol})$ in the de Rham
current cohomology \cite{preprint}.

Both choices (\ref{gf51}) and (\ref{gf52}) of $\Psi_5$ do not depend,
by construction,
on the local degrees of freedom of the fields. Moreover, the observables
$\sigma_n=\gamma_0^n$ are independent of sector of
${\cal B}_{gauge-fixing}$ that implements the constraint on the moduli
space.
After integrating over the
global degrees of freedom $\nu$-$\hat \nu$ of $\bar\psi$ and $A$, one
can also integrate over the local degrees of freedom of the
constraining sector
($\lambda$, $\bar\Gamma$, $\chi$, $\xi$, $b_z$, $c$, $\beta_z$, $\gamma$).
When zero modes are suitably taken into account,
such integrations give a net unit factor, since fermionic and bosonic
determinants compensate, as it is common in topological field theory.
The surviving fields are precisely those of the Verlinde
and Verlinde model and the only remnant of the constraining procedure
is the insertion of $c_g({\cal E}_{hol})$. Thus, the physical amplitudes
are
\begin{equation}
<\prod_{i=1}^s\sigma_{d_i}(x_i)>=\int_{{\cal M}_{g,s}}
c_g({\cal E}_{hol})\prod_{i=1}^s [c_1({\cal L}_i)]^{d_i},
\label{corre}
\end{equation}
as claimed in \cite{preprint}. ${\cal M}_{g,s}$ is
the moduli space of Riemann
surfaces $\Sigma_g$ of genus $g$ and $s$ marked points, while
$c_1({\cal L}_i)$ are the Mumford-Morita classes \cite{mumford}.
The selection rule is clearly
\begin{equation}
\sum_{i=1}^sd_i=2g-3+s.
\label{selection}
\end{equation}

Notice that, after the above integrations, the problem of the
difference
between true and formal dimensions \cite{preprint} is bypassed.
Neverthelsss, the fact that the right hand side of (\ref{selection})
contains $2g-3$ instead of something proportional to $g-1$ (and so to
the curvature) seems intriguing, since it is not straightforward to
repeat the Verlinde and Verlinde analysis of contact terms.
This is, to our opinion, a challenging feature of constrained
topological gravity.

In the case of the sphere the correlation functions are the same as in
ordinary topological gravity. For $g=1$, on the other hand, $\Omega=\tau$
and $c_1({\cal E}_{hol})={d\tau\wedge d\bar \tau\over (\tau-\bar \tau)^2}$,
which is the Poincar\'e metric. Its Poincar\'e dual is a point, that can be
chosen
at infinity. This corresponds to a torus with a pinched cycle
or, equivalently, a sphere with two identified points.
For $s=1$, (\ref{selection}) is zero, so that the moduli space is a point and
we can write
\begin{equation}
<\sigma_0>={1\over 2},
\end{equation}
the one half being a symmetry factor due to the identification
of the two points.
This correlation function  is the analogue of
$<\sigma_0\sigma_0\sigma_0>$ in genus zero. In view of
the above remarks, it is natural to
expect that the correlation functions in genus one are a half
of the corresponding correlation functions in genus zero, namely
when $\sum_{i=1}^kn_i=k$, then
\begin{equation}
<\sigma_0\prod_{i=1}^k\sigma_{n_i}>=
{1\over 2}<\sigma_0\sigma_0\sigma_0\prod_{i=1}^k\sigma_{n_i}>=
{1\over 2}{k!\over \prod_{i=1}^kn_i!}.
\end{equation}

One can conceive several variants of (\ref{corre}), in which
correlation
functions are products of Mumford-Morita classes times a fixed moduli
space cocycle. For example, one could replace $c_g({\cal E}_{hol})$
with $c_{g-k}({\cal E}_{hol})$, $0<k<g$. However, the case that we
have considered is the one that deserves particular attention, firstly
because it is suggested by physics, secondly because only $c_g({\cal
E}_{hol})$ is expressible as a determinant and can be easily inserted
in a field theoretical model, thirdly because $c_{g-k}({\cal E}_{hol})$
is not meaningful for all genera, but only for $g\geq k$. (Anyway,
the fact that for $k=1$, the right hand side of (\ref{selection})
is $2g-2+s$ perhaps deserves attention).

To conclude, many open questions
still remain to be answered and lots of possible
applications should be investigated in the future.
The first question is whether the correlation
functions of constrained topological gravity satisfy any integrable
hierarchy. In other words, one would like to know if one can
generalize the Kontsevich contruction \cite{kontsevich},
by identifying the set of fat graphs that describe the gravitational
instantons in two dimensions and by finding the corresponding matrix model.

A very promising chapter, still to be open, concerns the possible couplings
of contrained topological gravity to topological matter. One should
generalize to this case the analysis done for standard topological
gravity. In particular, one should investigate the meaning of the
equivariance condition \cite{verlindesquare,eguchietal} on the
physical states and what are the possible matter representatives
for the gravitational observables \cite{losev}.
Moreover it would be very interesting to know what are the effects of
the moduli space constraint on the holomorphic anomaly
\cite{cecotti}.

Finally, one can also think of a generalization of the contraining
mechanism proposed in this letter, by studying different
gauge-fixing terms for the global degrees of freedom.
In particular, one can wonder whether
in the standard theory of topological gravity
coupled to matter the moduli space contraint possesses
a representation in some sort of ``matter picture'' similar
to the ones of \cite{eguchietal,losev}.


\vspace{24pt}

\begin{thebibliography}{99}

\bibitem{preprint} D.\ Anselmi, P.\ Fr\'e, L.\ Girardello and P.\ Soriani,
``Constrained Topological Gravity from Twisted N=2 Liouville Theory'',
preprint SISSA/ISAS 49/94/EP, IFUM 468/FT, hepth/9404109, April 1994.

\bibitem{verlindesquare}
E.\ Verlinde and H.\ Verlinde, Nucl.\ Phys.\ B348 (1991) 457.

\bibitem{imbimbo} C.M.\ Becchi, R.\ Collina and C.\ Imbimbo,
{\it A Functional and Lagrangian Formulation of
Two Dimensional Topological
Gravity}, hepth 9406096, June 94;
talks given at the Fubinifest, Torino, February 1994
(to appear in the Proceedings, World Scientific Publishing) and at
the Trieste Workshop on
String Theory, April 1994.

\bibitem{eguchietal} T.\ Eguchi, H.\ Kanno, Y.\ Yamada and S.K.\ Yang,
Phys.\ Lett.\ B 305 (1993) 235.

\bibitem{dhoker} M.\ Schiffer and D.C.\ Spencer, ``Functionals of
finite Riemann surfaces'', Princeton University Press, Princeton 1954;

E. D' Hoker and D. H.\ Phong, Rev.\ Mod.\ Phys.\ 60 (1988) 917.

\bibitem{griffithsharris} P.\ Griffiths and J.\ Harris, {\it Principles of
Algebraic Geometry}, a Wiley-Interscience publication, 1978, pp.\
409-414.

\bibitem{mumford} S.\ Morita, Invent.\ Math.\ 90 (1987) 551;

D.\ Mumford, ``Towards an enumerative geometry of the moduli space of
curves. Arithmetic and Geometry'', Progr.\ Math.\ 36 (1983) 271.

\bibitem{kontsevich} M.\ Kontsevich, Comm.\ Math.\ Phys.\ 147 (1992) 1.

\bibitem{losev} A.\ Losev, {\it Descendants constructed from matter
field in topological Landau-Ginzburg model coupled to topological
gravity}, preprint TPI-MINN-92/40-T, 1992.

\bibitem{cecotti} M.\ Bershadsky, S.\ Cecotti, S.\ Ooguri and C.\
Vafa, Nucl.\ Phys.\ B405 (1993) 279.

\end{thebibliography}
\end{document}